\documentclass[twocolumn,showpacs,preprintnumbers,amsmath,amssymb]{revtex4}

\usepackage{graphicx}
\usepackage{dcolumn}
\usepackage{bm}
\usepackage{color}
\begin{document}

\preprint{APS/123-QED}

\title{Field-Induced Quadrupolar Quantum Criticality in PrV$_2$Al$_{20}$}

\author{Yasuyuki Shimura$^{1}$}
\email{simu@issp.u-tokyo.ac.jp}
\author{Masaki Tsujimoto$^{1}$}
\author{Bin Zeng$^{2}$}
\author{Luis Balicas$^2$}
\author{Akito Sakai$^{1,3}$}
\author{Satoru Nakatsuji$^{1,4}$}
\email{satoru@issp.u-tokyo.ac.jp}

\affiliation{$^{1}$Institute for Solid State Physics, The University of Tokyo, Kashiwa, Chiba 277-8581, Japan \\
$^{2}$National High Magnetic Field Laboratory (NHMFL), Florida State University, Tallahassee, Florida 32310, USA \\
$^{3}$I. Physikalisches Institut, Georg-August-Universit\"{a}t G\"{o}ttingen, 37077 G\"{o}ttingen, Germany \\
$^{4}$PRESTO, Japan Science and Technology Agency (JST), 4-1-8 Honcho Kawaguchi, Saitama 332-0012, Japan
}

\date{\today}

\begin{abstract}
PrV$_2$Al$_{20}$ is the heavy fermion superconductor based on the cubic $\Gamma_3$ doublet that exhibits non-magnetic quadrupolar ordering below $\sim $ 0.6 K. Our magnetotransport study on PrV$_2$Al$_{20}$ reveals field-induced quadrupolar quantum criticality   at $\mu_0 H_{\rm c} \sim 11 $ T applied along the [111] direction.  Near the critical field $\mu _0 H_{\rm c}$ required to suppress the quadrupolar state, we find a marked enhancement of the resistivity $\rho (H, T)$, a divergent effective mass of quasiparticles and concomitant non-Fermi liquid (NFL) behavior (i.e. $\rho(T) \propto T^n$ with $n \leq 0.5$). We also observe the Shubnikov de Haas-effect above $\mu_0 H_{\rm c}$, indicating   
the enhanced effective mass $m/m_0 \sim 10$. This reveals the competition between the nonmagnetic Kondo effect and the intersite quadrupolar coupling, leading to the pronounced NFL behavior in an extensive region of $T$ and $\mu_0 H$ emerging from the quantum critical point. 

\end{abstract}

\maketitle
Quantum criticality in correlated electron systems has attracted significant attention because of the formation of novel quantum phases such as exotic superconductivity in the vicinity of a quantum critical point (QCP) \cite{Pines07}. Moreover, the breakdown of the standard Fermi-liquid behavior has been seen almost routinely nearby a magnetic QCP in a variety of strongly correlated electron systems, ranging from cuprates, iron pnictides and heavy fermion intermetallics \cite{QC_Lohneysen07, QC_Gegenwart08}. Whether another type of instability such as orbital ordering and its associated orbital fluctuations may drive novel types of metallic state and unconventional superconductivity has been an active area of research \cite{Pines07, PrOs4Sb12_Bauer02, PrOs4Sb12_Miyake03, Hattori10, Kontani10, Yanagi10}. Experimentally, however, quantum criticality due solely to an orbital origin has never been observed in metallic systems. 

For the study of quantum criticality, $4f$-electron systems are well suited and provide various archetypical examples due to the availability of high-purity single crystals and the relatively low characteristic energy scales which are highly tunable by disorder-free control parameters such as magnetic field and pressure. 
To date, among heavy-fermion intermetallics, most of the study on QC have been made for compounds containing either Ce ($4f^1$) or Yb ($4f^{13}$) (see, for example, Refs. \onlinecite{Mathur98, Hegger00_CeRhIn5, Paglione,YbRh2Si2_Gegenwart02, YbAlB4_Nakatsuji08}) ions whose crystalline-electric-field (CEF) ground-state is composed of Kramers doublets and therefore magnetic. 

In transition metal systems, the coupling between spins and orbitals is unavoidable and produces various interesting spin-orbital ordered and disordered states \cite{tokuranagaosa,BCSO}. In contrast, an $f$-electron system may provide a non-magnetic CEF ground-state doublet, where orbitals are the only active degree of freedom. In effect, some Pr ($4f^2$)-based cubic compounds are found to host the $\Gamma _3$ non-Kramers ground-state  doublet, which has no magnetic moment but carries an electric quadrupole moment. In these systems, due to strong intra-atomic spin-orbit coupling, the total angular momentum $J$ represents the magnetic and orbital states, and in particular, the electric quadrupole moment corresponds to the orbital degree of freedom. 
A number of cubic 4$f^2$ $\Gamma _3$ systems have been studied and various interesting electric phenomena have been experimentally reported including a ferro and antiferro quadrupolar ordering based on the RKKY-type interaction \cite{PrPb3_Morin82, PrPtBi_Suzuki97,PrPb3_Onimaru05, PrIr2Zn20_Onimaru10,PrIr2Zn20_Onimaru11}. As a competing effect, a nonmagnetic form of the Kondo effect is proposed that quenches quadrupole moments \cite{Cox87, PrInAg2_Yaster96}. Thus, the tuning of these competing effects may lead to quadrupolar QC. 

In fact, quadrupolar quantum criticality has been suggested by recent experiments on the new cubic $\Gamma_3$ systems Pr$T_2$Al$_{20}$, where $T$ corresponds to a transition metal such as Ti and V \cite{PrTr2Al20_Sakai11, PrTi2Al20_Sakai12,PrTi2Al20_Matsubayashi12,PrV2Al20_Tsujimoto14}. In these systems, the hybridization between the $f$-moments and the conduction ($c-$) electrons is found to be not only strong but tunable.
First, the strong $c$-$f$ hybridization is evident from a number of observations, including the Kondo-effect in the resistivity (i.e. $\rho(T) \propto -\ln T$)\cite{PrTr2Al20_Sakai11}, a Kondo-resonance peak observed near the Fermi-energy \cite{PrTi2Al20_Matsunami11}, and a large hyperfine constant in the NMR measurements \cite{PrTi2Al20_Tokunaga13}. 
Secondly, the tunability of the hybridization strength in Pr$T_2$Al$_{20}$ is demonstrated by both chemical and physical pressure measurements. The substitution of Ti by V enhances the Kondo-effect and induces an anomalous metallic behavior due to the hybridization. PrTi$_2$Al$_{20}$ exhibits a ferro-quadrupole ordering at $T_{\rm Q} = $ 2 K with a subsequent superconducting (SC) transition at $T_{\rm c} = $ 0.2 K \cite{PrTr2Al20_Sakai11, PrTi2Al20_Sakai12}. While the SC effective mass of PrTi$_2$Al$_{20}$ is moderately enhanced under ambient pressure, i.e. $m^*/m_0 \sim 16$ ($m_0$ is the free electron mass), the application of pressure increases $T_c$ up to 1 K and $m^*/m_0$ up to 110 at $P \sim 8$ GPa, while suppressing $T_{\rm Q}$ \cite{PrTi2Al20_Matsubayashi12}.This indicates that the pressure-induced heavy-fermion superconductivity emerges possibly in the vicinity of a putative quadrupolar QCP.


Evidence for strong hybridization in PrV$_2$Al$_{20}$ is further provided by the recent discovery of heavy-fermion superconductivity at $T_c = 50$ mK with a large specific heat jump $\Delta C/T \sim 0.3$ J/mol K$^2$ below $T_{\rm Q} = 0.6 - 0.7$ K under ambient pressure \cite{PrV2Al20_Tsujimoto14}.
The effective mass of the quasiparticles participating into the superconducting condensate is found to be as large as 140 $m_0$, one order of magnitude larger than that of its Ti analog \cite{PrTi2Al20_Sakai12}. This result indicates PrV$_2$Al$_{20}$ should be located in the vicinity of a QCP associated \emph{only} with multipole moments.

To realize such quantum criticality due to the multipole moments, the magnetic field, in addition to pressure, is another useful control parameter that couples quadratically with quadrupole moments, thus more weakly than with magnetic moments. 
The high-field phase diagram  in PrV$_2$Al$_{20}$ was investigated through specific-heat measurements under fields up to 9 T applied along all three main crystallographic orientations. Overall, the low-field phase boundaries for the [100], [110] and [111] directions are very similar to one another and nearly independent of $\mu_0 H$, as often observed in various quadrupolar ordered systems \cite{PrTr2Al20_Sakai11, PrIr2Zn20_Onimaru11}. Moreover, the high field magnetization measurement revealed a field induced first-order transition at $\mu_0 H \sim 11$ T for $H \| [100]$ which is most likely due to the switching of the quadrupole order parameter in the $\Gamma _3$ ground doublet \cite{PrV2Al20_Shimura13}. 

Here, we report the discovery of field-tuned quantum criticality purely based on the quadrupolar (orbital) degrees of freedom at ambient pressure in PrV$_2$Al$_{20}$.
We have studied through magneto-transport measurements, the magnetic phase diagram of PrV$_2$Al$_{20}$ for $H || [111]$, and found unusual non-Fermi liquid behavior with $\rho = \rho_0 + AT^n$ with $n \leq 0.5$, a divergent effective mass of quasiparticle and a large enhancement in the residual resistivity $\rho_0$ around the magnetic field-induced quantum-phase transition at the critical-field $\mu_0 H_{\rm c} \sim 11$ T, where the quadrupolar transition temperature is suppressed to absolute zero. In addition, our observation of quantum oscillation reveals a heavy mass state with $m^*/m_0 > 10$ in the paraquadrupolar state beyond  $\mu_0 H_{\rm c}$, indicating nonmagnetic Kondo effect competing with quadrupolar coupling as the origin of the pronounced quantum criticality observed over an extensive region of $T$ and $\mu_0 H$. The experimental condition is described in detail in the supplemental material \cite{SM}.

\begin{figure}[t]
\begin{center}
\includegraphics[width=90mm]{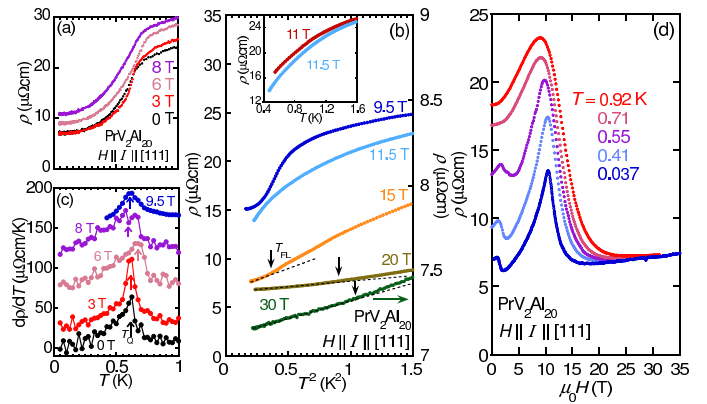}
\end{center}
\caption{(Color online) (a, b) Longitudinal resistivity $\rho$ for PrV$_2$Al$_{20}$ single-crystals as a function of temperature $T$ under magnetic fields. The solid arrow indicates the characteristic temperature $T_{\rm FL}$ below which $\rho (T)$ exhibits a $T^2$ dependence (broken line). (c) $T$ dependence of the temperature derivative of the resistivity, $d \rho/d T$ under field below 10 T. The data points are vertically shifted for clarity. The arrows indicate the quadrupolar transition temperature $T_{\rm Q}$.
(d) $\rho$ as a function of $\mu_0 H$ $ \| [111]$ at various temperatures.
}
\label{rtrh}
\end{figure}

The temperature dependence of the resistivity $\rho (T)$ in both the low field ($\mu_0 H \leq 8$ T) and the high field ($\mu_0 H \geq 9.5$ T) regions  are presented in Figs. 1(a) and 1(b), respectively. 
Both magnetic field and electric current were applied along the [111] direction.
When the field is lower than 10 T, one observes a sudden decrease in $\rho (T)$ upon cooling due to the quadrupolar phase-transition. Correspondingly, a peak is observed in the $T$ derivative of the resistivity at a low temperature $T_{\rm Q}$ which systematically changes with field, as shown in Fig. 1(c).
On the other hand, the resistivity under fields surpassing 11 T shows such a smooth temperature dependence that the anomaly associated with the quadrupolar ordering no longer exists. Thus, a quantum phase transition between the low-field quadrupolar and high-field paraquadrupolar phases should be located at $\mu_0 H_{\rm c} \sim $ 11 T. Such a field-indcued suppression of a quadrupolar phase was predicted for the Pr$T_2$Al$_{20}$ system using a mean-field theory based on a localized picture \cite{G3_Hattori14}.

Remarkably, under $\mu_0 H = 11$ T, $\rho (T)$ exhibits a sublinear $T$ dependence, in sharp contrast to the Fermi-liquid behavior $\rho (T) \propto T^2$, while a super-linear $T$ dependence, which proportional to the $T^2$ law, appears at higher fields (Fig. 2(b)).
A detailed analysis of the field and temperature dependence of $\rho$ will be described below.
Figure 1(d) displays the field dependence of the magnetoresistivity for $T < 1$ K. 
$\rho (H)$ exhibits a marked peak at $\mu_0 H_{\rm c} \sim $ 11 T, in addition to the sharp decrease as a function of  temperature due to the quadrupolar ordering at $\sim 0$ T. 
If this peak resulted solely from thermal critical fluctuations associated with a finite temperature transition between the quadrupolar phase to the paraquadrupolar state, this peak would be expected to shift to lower fields with increasing $T$ and eventually reach zero field near $T_{\rm Q} \sim 0.6$ K.
However in PrV$_2$Al$_{20}$, a pronounced peak is still observed at nearly the same field $\sim $ 11 T at $T \sim 1$ K $>>$ $ T_{\rm Q}$, indicating the development of quantum critical scattering at $\sim \mu_0 H_{\rm c} $ at much higher $T$ than $T_{\rm Q}$.
In addition to the main peak at $\sim 11 $~T, a small kink was observed at $\sim$ 3~T in the quadrupolar ordered state below 0.6~K.
This may be associated with the change in the order parameters such as the lifting of the degeneracy between the O$^2_0$ and O$^2_2$ states \cite{G3_Hattori14}.

\begin{figure}[t]
\begin{center}
\includegraphics[width=90mm]{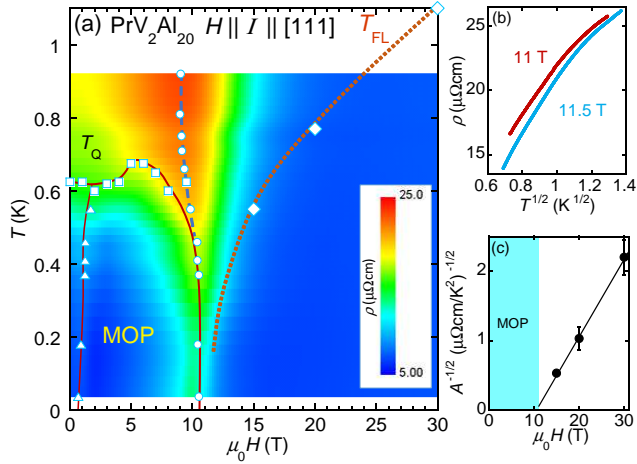}
\end{center}
\caption{(Color online) (a) Magnetic phase diagram of PrV$_2$Al$_{20}$ for fields, and current $I$, parallel to the [111] direction. 
The color plot indicates the $\rho(T,H)$ values obtained from $H$ scans under constant $T$.
Circles indicate the peak position at $\mu_0 H_{\rm c} \sim 11$ T, separating the low-field multipole ordering phase (MOP) and the paraquadrupolar state at high fields. Triangles represent anomalies existing below $T_{\rm Q}$ at $ \sim 3$ T observed in $\rho (H)$. The solid line and broken line show the transition temperature/field of the multipole ordered phase and the peak position in $\rho (H)$, respectively.
Squares and diamonds respectively indicate $T_{\rm Q}$ determined from $\rho(T)$ and a characteristic temperature $T_{\text{FL}}$ below which $\rho (T)$ follows the Fermi-liquid $T^2$ law.
The dotted line is a guide to the eyes.
(b) $\rho$ as a function of $T^{1/2}$ under $\mu_0 H = 11$ and 11.5 T.
(c) Field dependence of $A^{-1/2}$, where $A$ is the $T^2$ coefficients in $\rho (T)$.
The solid line is the fit to $\sqrt{A} = \sqrt{A_0}/(\mu_0 H - \mu_0 H _{\rm c})$.
}
\label{phase}
\end{figure}

To clearly illustrate the field-induced enhancement of the resistivity above $T_{\rm Q}$, Figure 2(a) shows a contour plot of $\rho$ as a function of both $T$ and $H \parallel [111]$.
The distinct color change between blue and green at $\sim $ 0.5 K found below 10 T follows the quadrupolar transition temperature $T_{\rm Q}$ (square) determined by the temperature scans discussed above. This line of $T_{\rm Q}$ connects smoothly with the line depicting the peak position (circle) in the field dependence of the resistivity $\rho (H)$. These results indicate that the peak in the magnetoresistance at low temperatures corresponds to the quadrupolar phase boundary (solid line in Fig. 2(a)), which reaches the quantum critical point at $ \mu_0 H_{\rm c} \sim 11$ T. 

Remarkably, the peak observed in $\rho (H)$ around 11 T survives up to $ \sim 1~{\rm K} > T_{\rm Q}$ as discussed above, and this pronounced peak in $\rho(T,H)$ (red region in Fig. 2(a)), which is observed in the paraquadrupolar regime cannot be explained within a simple localized $f$-moment scenario. 
Figure 2 (b) shows $\rho (T)$ as a function of $T^{1/2}$ at $\mu_0 H = 11$ T near the quantum-critical field. 
$\rho (T)$ under $\mu_0 H = 11 $ and 11.5 T shows a concave curvature, indicating that the exponent $n$ in $\rho (T) = \rho _0 + AT^n$ is even smaller than $0.5$.
On the contrary, above 15 T, $\rho (T)$ exhibits a convex curvature indicating the emergence of Fermi liquid behavior at the lowest $T$s, as shown in Fig. 1 (b).
The characteristic temperature $T_{\rm FL}$ below which $\rho (T)$ displays FL-behavior or $\rho (T) = \rho _0 + AT^2$ increases with field (Fig. 2 (a)).
Figure 2 (c) indicates the field dependence of $A^{-1/2}$ obtained from $\rho (T)$ above 15 T. 
Accordingly, upon approaching the quantum-critical field, the corresponding $A$ values diverge, exceeding $ \sim 5 \ \mu \Omega \rm{cm}/\rm{K}^2$. It can be fit to $\sqrt{A} = \sqrt{A_0}/(\mu_0 H - \mu_0 H _{\rm c}^{A})$ with $\sqrt{A_0} = 8.92$ $(\mu \Omega $cm T/K$^2)^{1/2}$, and $\mu_0 H_{\rm c}^{A} = 10.5$ T (Fig. 2(c)). Significantly, $\mu_0 H_{\rm c}^{A}$ is found consistent with the critical field $\sim 11$ T determined by the peak in $\rho(H)$.
According to the Kadowaki-Woods relation, the critical enhancement in $A^{1/2}$ indicates the divergence of the effective mass upon approaching the QCP.

The evolution of the crystal electric field (CEF) scheme under field alone cannot explain the quantum criticality. The CEF scheme was determined by previous detailed magnetization measurements at low temperatures \cite{PrV2Al20_Proc_Araki14}. 
We simulated the effect of the magnetic field on this CEF scheme for fields up to 30 T (see the supplemental material) \cite{SM}. 
Under a field, the $\Gamma _3$  ground doublet opens a gap because of the van-Vleck type mixing of the excited magnetic triplet states. Our CEF calculation indicates that the gap is found to increase smoothly with the magnetic field without any anomaly at $H_{\rm c}$ and thus does not explain the peak in $\rho (H)$.
In addition, the gap of the ground doublet is found to be approximately 2 K at $H_{\rm c}$.
In contrast, at zero field below $\sim 20$ K, PrV$_2$Al$_{20}$ exhibits anomalous metallic behavior such as $T^{1/2}$ dependence of the susceptibility, and $T^n$ dependence of the resistivity with $n \le 0.5$ \cite{PrTr2Al20_Sakai11,PrV2Al20_Tsujimoto14,PrV2Al20_Shimura13}. This indicates that the characteristic scale for the nonmagnetic Kondo effect based on the hybridization between the $\Gamma_3$ doublet and the conduction electrons is more than 20 K. 
Therefore, the nonmagnetic Kondo effect has a much higher energy scale than the CEF gap at the critical field, hence the quantum critical anomaly near $\mu_0 H_{\rm c}$ should mainly come from the competition between the intersite quadrupolar coupling, and the nonmagnetic Kondo effect.

In fact, the sharp magnetoresistance peak at $\mu_0 H_{\rm c} \sim $ 11 T, which is observed even above $T_{\rm Q}$, supports a significant role for hybridization effects in the quantum criticality.
In effect, at a pressure-induced quantum critical point, the enhancement in the residual resistivity has been reported and attributed to quantum critical fluctuations \cite{QC_Lohneysen07}.
In the case of Pr-based compounds with the $\Gamma _3$ ground doublet, such an enhancement in $\rho_0$ was also observed under zero field, especially above $\sim 7$ GPa in PrTi$_2$Al$_{20}$, which is accompanied by the suppression of the quadrupole-order \cite{PrTi2Al20_Matsubayashi12}.
$T_{\rm c}$ as well as the effective mass $m^*$ also increase considerably above $\sim 7$ GPa, while $T_{\rm Q}$ starts to decrease, suggesting the proximity to a putative quantum critical point \cite{PrTi2Al20_Matsubayashi12}.
In PrV$_2$Al$_{20}$, the similarly dramatic enhancement in both $\rho_0$ and in $m^* \sim A^{1/2}$ under magnetic field, coupled to the anomalous $T$-dependence of the resistivity at the critical field $\mu_0 H_{\rm c} \simeq 11$ T,  provides firm experimental evidence for field-induced quantum criticality based on the strong hybridization between the conduction electrons and the nonmagnetic quadrapolar / orbital degrees of freedom in the $\Gamma_3$ ground doublet. 
As discussed above, the resistivity follows $\rho = \rho_0 + AT^n$, with $n \leq 0.5$ at $\mu_0 H_{\rm c}$, in sharp contrast to $n  = 1$ and 1.5,  which are usually observed around a QCP in Ce/Yb based heavy-fermion compounds with a ground Kramers  doublet \cite{Mathur98, YbRh2Si2_Gegenwart02, Hegger00_CeRhIn5,Paglione,YbAlB4_Nakatsuji08}. Such anomalous temperature dependence with $n \le 0.5$ should be specific to the quantum critical phenomena associated with quadrupolar degrees of freedom.

\begin{figure}[t]
\begin{center}
\includegraphics[width=90mm]{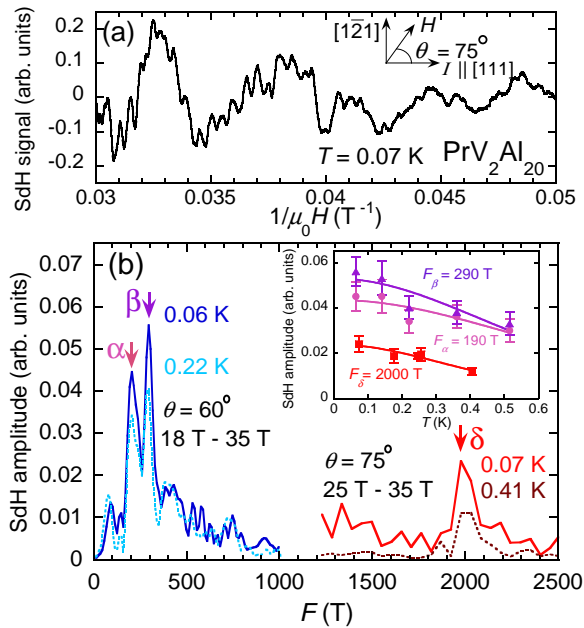}
\end{center}
\caption{(Color online) (a) Typical trace of the oscillatory signal due to the Shubnikov de Haas-effect (SdH), superimposed onto the $\rho(H)$ trace as a function of $\mu_0 H^{-1}$ for $T = 70$ mK and for an angle $\theta = 75^{\circ}$ from [111] direction. (b) Fast Fourier transform (FFT) of the oscillatory signal for $\theta = 60^{\circ}$ and at $T= 60$ and 220 mK, respectively. One observes two prominent peaks at $F_{\alpha}\sim 190$ T and $F_{\beta} \sim 290$ T. The same figure shows the FFT spectra for $\theta = 75^{\circ}$ and $T = 70$ and 410 mK. For this orientation one detects a higher frequency $F_{\delta} \sim 2$ kT. Inset: Lifshitz-Kosevich fits of the SdH amplitude as a function of $T$ from which we extract the corresponding effective masses $m^{\star}$.}
\label{sdh}
\end{figure}

Finally, the high-quality of our single crystals allows us to observe the Shubnikov de Haas (SdH)-effect above the critical field. Figure 3(a) illustrates a representative trace of the oscillatory signal (after the subtraction of a polynomial fit) at $T = 70$ mK as a function of inverse field $1/\mu_0 H$ superimposed into $\rho(H)$, under fields aligned at an angle $\theta = 75^{\circ}$ from the [111] direction and beyond the critical value required to suppress the quadrupolar state. The corresponding fast Fourier-transform spectra is shown in Fig. 3 (b) for two values of $T$, respectively at 60 mK (dark blue trace) and 220 mK (clear blue trace). One detects two main small frequencies $F_{\alpha} \sim 190$ T and $F_{\beta} \sim 290$ T. Figure 3 (b) also displays two other traces for $H$ at an angle $\theta = 75^{\circ}$, and $T= 70$ mK (dark orange trace) and 410 mK (brown trace) respectively, revealing a higher frequency $F_{\delta} \sim 2$ kT.

An important piece of information is provided by the inset in Fig. 3 (b) which displays the amplitude of the FFT peaks as a function of $T$: solid lines are fits to the Lifshitz-Kosevich thermal damping term, i.e. $X/\sinh X$ with $X = 4 \pi^3 k_B m^{\star}T/eH$ (where $k_B$ is the Boltzmann constant, and $e$ is the electron charge) from which we extract the effective mass $m^{\star}$ for each $F$. The resulting mass values are $m^{\star}_{\alpha} = (5.7 \pm 1.2)m_0$, $m^{\star}_{\beta} = (6.8 \pm 1.2)m_0$ and $m^{\star}_{\delta} = (10.6 \pm 1.2)m_0$, which are still moderately heavy, indicating the presence of the nonmagnetic Kondo effect based on the $c-f$ hybridization. This is also consistent with the fact that the nonmagnetic Kondo effect has a much higher scale than the CEF gap at the critical field. Thus, the quantum-critical behavior unveiled here cannot be ascribed to an $f$-electron localization transition involving the suppression of the Kondo-effect at low fields. The charge carriers are so  heavy as $m^* \sim 140 \ m_0$ at zero-field, and still display moderately heavy masses at high fields. In contrast with the divergence found in the $A$ coefficient as $\mu_0 H \rightarrow \mu_0 H_{\rm c}$, we could not find evolution of the cyclotron masses upon approaching the QCP thus indicating that other, probably undetected, Fermi surface sheets are involved in the QC phenomenology displayed by PrV$_2$Al$_{20}$, and that the detected cyclotronic orbits remain oblivious to the quantum fluctuations associated with orbital degrees of freedom. The observation of the SdH signal paves the path to further clarify the electronic structure and to understand the strong screening effects leading to the quadrupolar quantum criticality involving the prominent non-Fermi liquid behavior in PrV$_2$Al$_{20}$. \nocite{LLW, PrFe4P12_Aoki}


\begin{acknowledgments}
We thank Y. Matsumoto, T. Tomita, T. Sakakibara, K. Araki. Y. Uwatoko, K. Matsubayashi, J. Suzuki, K. Miyake, K. Hattori and H. Kusunose for useful discussions.
This work was partially supported by Grants-in-Aid (No. 25707030 and 25887015) from JSPS, and by PRESTO, JST, Japan.
Y.~S. is partially supported by ICAM.
The NHMFL is supported by NSF through NSF-DMR-0084173 and the
State of Florida. L.~B. is supported by DOE-BES through award DE-SC0002613.
This work was supported in part by NSF Grant No. PHYS-1066293 and we acknowledge the hospitality of the Aspen Center for Physics.
\end{acknowledgments}


\begin{thebibliography}{34}
\expandafter\ifx\csname natexlab\endcsname\relax\def\natexlab#1{#1}\fi
\expandafter\ifx\csname bibnamefont\endcsname\relax
  \def\bibnamefont#1{#1}\fi
\expandafter\ifx\csname bibfnamefont\endcsname\relax
  \def\bibfnamefont#1{#1}\fi
\expandafter\ifx\csname citenamefont\endcsname\relax
  \def\citenamefont#1{#1}\fi
\expandafter\ifx\csname url\endcsname\relax
  \def\url#1{\texttt{#1}}\fi
\expandafter\ifx\csname urlprefix\endcsname\relax\def\urlprefix{URL }\fi
\providecommand{\bibinfo}[2]{#2}
\providecommand{\eprint}[2][]{\url{#2}}

\bibitem[{\citenamefont{Monthoux et~al.}(2007)\citenamefont{Monthoux, Pines,
  and Lonzarich}}]{Pines07}
\bibinfo{author}{\bibfnamefont{P.}~\bibnamefont{Monthoux}},
  \bibinfo{author}{\bibfnamefont{D.}~\bibnamefont{Pines}}, \bibnamefont{and}
  \bibinfo{author}{\bibfnamefont{G.~G.} \bibnamefont{Lonzarich}},
  \bibinfo{journal}{Nature} \textbf{\bibinfo{volume}{450}},
  \bibinfo{pages}{1177} (\bibinfo{year}{2007}).

\bibitem[{\citenamefont{L{\"o}hneysen et~al.}(2007)\citenamefont{L{\"o}hneysen,
  Rosch, Vojta, and W{\"o}lfle}}]{QC_Lohneysen07}
\bibinfo{author}{\bibfnamefont{H.~V.} \bibnamefont{L{\"o}hneysen}},
  \bibinfo{author}{\bibfnamefont{A.}~\bibnamefont{Rosch}},
  \bibinfo{author}{\bibfnamefont{M.}~\bibnamefont{Vojta}}, \bibnamefont{and}
  \bibinfo{author}{\bibfnamefont{P.}~\bibnamefont{W{\"o}lfle}},
  \bibinfo{journal}{Rev. Mod. Phys.} \textbf{\bibinfo{volume}{79}},
  \bibinfo{pages}{1015} (\bibinfo{year}{2007}).

\bibitem[{\citenamefont{Gegenwart et~al.}(2008)\citenamefont{Gegenwart, Si, and
  Steglich}}]{QC_Gegenwart08}
\bibinfo{author}{\bibfnamefont{P.}~\bibnamefont{Gegenwart}},
  \bibinfo{author}{\bibfnamefont{Q.}~\bibnamefont{Si}}, \bibnamefont{and}
  \bibinfo{author}{\bibfnamefont{F.}~\bibnamefont{Steglich}},
  \bibinfo{journal}{Nature Phys.} \textbf{\bibinfo{volume}{4}},
  \bibinfo{pages}{186} (\bibinfo{year}{2008}).

\bibitem[{\citenamefont{Bauer et~al.}(2002)\citenamefont{Bauer, Frederick, Ho,
  Zapf, and Maple}}]{PrOs4Sb12_Bauer02}
\bibinfo{author}{\bibfnamefont{E.~D.} \bibnamefont{Bauer}},
  \bibinfo{author}{\bibfnamefont{N.~A.} \bibnamefont{Frederick}},
  \bibinfo{author}{\bibfnamefont{P.-C.} \bibnamefont{Ho}},
  \bibinfo{author}{\bibfnamefont{V.~S.} \bibnamefont{Zapf}}, \bibnamefont{and}
  \bibinfo{author}{\bibfnamefont{M.~B.} \bibnamefont{Maple}},
  \bibinfo{journal}{Phys. Rev. B} \textbf{\bibinfo{volume}{65}},
  \bibinfo{pages}{100506} (\bibinfo{year}{2002}).

\bibitem[{\citenamefont{Miyake et~al.}(2003)\citenamefont{Miyake, Kohno, and
  Harima}}]{PrOs4Sb12_Miyake03}
\bibinfo{author}{\bibfnamefont{K.}~\bibnamefont{Miyake}},
  \bibinfo{author}{\bibfnamefont{H.}~\bibnamefont{Kohno}}, \bibnamefont{and}
  \bibinfo{author}{\bibfnamefont{H.}~\bibnamefont{Harima}},
  \bibinfo{journal}{J. Phys.: Condensed Matter} \textbf{\bibinfo{volume}{15}},
  \bibinfo{pages}{L275} (\bibinfo{year}{2003}).

\bibitem[{\citenamefont{Hattori}(2010)}]{Hattori10}
\bibinfo{author}{\bibfnamefont{K.}~\bibnamefont{Hattori}}, \bibinfo{journal}{J.
  Phys. Soc. Jpn.} \textbf{\bibinfo{volume}{79}}, \bibinfo{pages}{114717}
  (\bibinfo{year}{2010}).

\bibitem[{\citenamefont{Kontani and Onari}(2010)}]{Kontani10}
\bibinfo{author}{\bibfnamefont{H.}~\bibnamefont{Kontani}} \bibnamefont{and}
  \bibinfo{author}{\bibfnamefont{S.}~\bibnamefont{Onari}},
  \bibinfo{journal}{Phys. Rev. Lett.} \textbf{\bibinfo{volume}{104}},
  \bibinfo{pages}{157001} (\bibinfo{year}{2010}).

\bibitem[{\citenamefont{Yanagi et~al.}(2010)\citenamefont{Yanagi, Yamakawa,
  Adachi, and Ono}}]{Yanagi10}
\bibinfo{author}{\bibfnamefont{Y.}~\bibnamefont{Yanagi}},
  \bibinfo{author}{\bibfnamefont{Y.}~\bibnamefont{Yamakawa}},
  \bibinfo{author}{\bibfnamefont{N.}~\bibnamefont{Adachi}}, \bibnamefont{and}
  \bibinfo{author}{\bibfnamefont{Y.}~\bibnamefont{Ono}}, \bibinfo{journal}{J.
  Phys. Soc. Jpn.} \textbf{\bibinfo{volume}{79}}, \bibinfo{pages}{123707}
  (\bibinfo{year}{2010}).

\bibitem[{\citenamefont{Mathur et~al.}(1998)\citenamefont{Mathur, Grosche,
  Julian, Walker, Freye, Haselwimmer, and Lonzarich}}]{Mathur98}
\bibinfo{author}{\bibfnamefont{N.~D.} \bibnamefont{Mathur}},
  \bibinfo{author}{\bibfnamefont{F.~M.} \bibnamefont{Grosche}},
  \bibinfo{author}{\bibfnamefont{S.~R.} \bibnamefont{Julian}},
  \bibinfo{author}{\bibfnamefont{I.~R.} \bibnamefont{Walker}},
  \bibinfo{author}{\bibfnamefont{D.~M.} \bibnamefont{Freye}},
  \bibinfo{author}{\bibfnamefont{R.~K.~W.} \bibnamefont{Haselwimmer}},
  \bibnamefont{and} \bibinfo{author}{\bibfnamefont{G.~G.}
  \bibnamefont{Lonzarich}}, \bibinfo{journal}{Nature}
  \textbf{\bibinfo{volume}{394}}, \bibinfo{pages}{39} (\bibinfo{year}{1998}).

\bibitem[{\citenamefont{Hegger et~al.}(2000)\citenamefont{Hegger, Petrovic,
  Moshopoulou, Hundley, Sarrao, Fisk, and Thompson}}]{Hegger00_CeRhIn5}
\bibinfo{author}{\bibfnamefont{H.}~\bibnamefont{Hegger}},
  \bibinfo{author}{\bibfnamefont{C.}~\bibnamefont{Petrovic}},
  \bibinfo{author}{\bibfnamefont{E.~G.} \bibnamefont{Moshopoulou}},
  \bibinfo{author}{\bibfnamefont{M.~F.} \bibnamefont{Hundley}},
  \bibinfo{author}{\bibfnamefont{J.~L.} \bibnamefont{Sarrao}},
  \bibinfo{author}{\bibfnamefont{Z.}~\bibnamefont{Fisk}}, \bibnamefont{and}
  \bibinfo{author}{\bibfnamefont{J.~D.} \bibnamefont{Thompson}},
  \bibinfo{journal}{Phys. Rev. Lett.} \textbf{\bibinfo{volume}{84}},
  \bibinfo{pages}{4986} (\bibinfo{year}{2000}).

\bibitem[{\citenamefont{Paglione et~al.}(2003)\citenamefont{Paglione, Tanatar,
  Hawthorn, Boaknin, Hill, Ronning, Sutherland, Taillefer, Petrovic, and
  Canfield}}]{Paglione}
\bibinfo{author}{\bibfnamefont{J.}~\bibnamefont{Paglione}},
  \bibinfo{author}{\bibfnamefont{M.~A.} \bibnamefont{Tanatar}},
  \bibinfo{author}{\bibfnamefont{D.~G.} \bibnamefont{Hawthorn}},
  \bibinfo{author}{\bibfnamefont{E.}~\bibnamefont{Boaknin}},
  \bibinfo{author}{\bibfnamefont{R.~W.} \bibnamefont{Hill}},
  \bibinfo{author}{\bibfnamefont{F.}~\bibnamefont{Ronning}},
  \bibinfo{author}{\bibfnamefont{M.}~\bibnamefont{Sutherland}},
  \bibinfo{author}{\bibfnamefont{L.}~\bibnamefont{Taillefer}},
  \bibinfo{author}{\bibfnamefont{C.}~\bibnamefont{Petrovic}}, \bibnamefont{and}
  \bibinfo{author}{\bibfnamefont{P.~C.} \bibnamefont{Canfield}},
  \bibinfo{journal}{Phys. Rev. Lett.} \textbf{\bibinfo{volume}{91}},
  \bibinfo{pages}{246405} (\bibinfo{year}{2003}).

\bibitem[{\citenamefont{Gegenwart et~al.}(2002)\citenamefont{Gegenwart,
  Custers, Geibel, Neumaier, Tayama, Tenya, Trovarelli, and
  Steglich}}]{YbRh2Si2_Gegenwart02}
\bibinfo{author}{\bibfnamefont{P.}~\bibnamefont{Gegenwart}},
  \bibinfo{author}{\bibfnamefont{J.}~\bibnamefont{Custers}},
  \bibinfo{author}{\bibfnamefont{C.}~\bibnamefont{Geibel}},
  \bibinfo{author}{\bibfnamefont{K.}~\bibnamefont{Neumaier}},
  \bibinfo{author}{\bibfnamefont{T.}~\bibnamefont{Tayama}},
  \bibinfo{author}{\bibfnamefont{K.}~\bibnamefont{Tenya}},
  \bibinfo{author}{\bibfnamefont{O.}~\bibnamefont{Trovarelli}},
  \bibnamefont{and} \bibinfo{author}{\bibfnamefont{F.}~\bibnamefont{Steglich}},
  \bibinfo{journal}{Phys. Rev. Lett.} \textbf{\bibinfo{volume}{89}},
  \bibinfo{pages}{056402} (\bibinfo{year}{2002}).

\bibitem[{\citenamefont{Nakatsuji et~al.}(2008)\citenamefont{Nakatsuji, Kuga,
  Machida, Tayama, Sakakibara, Karaki, Ishimoto, Yonezawa, Maeno, Pearson
  et~al.}}]{YbAlB4_Nakatsuji08}
\bibinfo{author}{\bibfnamefont{S.}~\bibnamefont{Nakatsuji}},
  \bibinfo{author}{\bibfnamefont{K.}~\bibnamefont{Kuga}},
  \bibinfo{author}{\bibfnamefont{Y.}~\bibnamefont{Machida}},
  \bibinfo{author}{\bibfnamefont{T.}~\bibnamefont{Tayama}},
  \bibinfo{author}{\bibfnamefont{T.}~\bibnamefont{Sakakibara}},
  \bibinfo{author}{\bibfnamefont{Y.}~\bibnamefont{Karaki}},
  \bibinfo{author}{\bibfnamefont{H.}~\bibnamefont{Ishimoto}},
  \bibinfo{author}{\bibfnamefont{S.}~\bibnamefont{Yonezawa}},
  \bibinfo{author}{\bibfnamefont{Y.}~\bibnamefont{Maeno}},
  \bibinfo{author}{\bibfnamefont{E.}~\bibnamefont{Pearson}},
  \bibnamefont{et~al.}, \bibinfo{journal}{Nature Phys.}
  \textbf{\bibinfo{volume}{4}}, \bibinfo{pages}{603} (\bibinfo{year}{2008}).

\bibitem[{\citenamefont{Tokura and Nagaosa}(2000)}]{tokuranagaosa}
\bibinfo{author}{\bibfnamefont{Y.}~\bibnamefont{Tokura}} \bibnamefont{and}
  \bibinfo{author}{\bibfnamefont{N.}~\bibnamefont{Nagaosa}},
  \bibinfo{journal}{Science} \textbf{\bibinfo{volume}{288}},
  \bibinfo{pages}{462} (\bibinfo{year}{2000}).

\bibitem[{\citenamefont{Nakatsuji et~al.}(2012)\citenamefont{Nakatsuji, Kuga,
  Kimura, Satake, Katayama, Nishibori, Sawa, Ishii, Hagiwara, Bridges
  et~al.}}]{BCSO}
\bibinfo{author}{\bibfnamefont{S.}~\bibnamefont{Nakatsuji}},
  \bibinfo{author}{\bibfnamefont{K.}~\bibnamefont{Kuga}},
  \bibinfo{author}{\bibfnamefont{K.}~\bibnamefont{Kimura}},
  \bibinfo{author}{\bibfnamefont{R.}~\bibnamefont{Satake}},
  \bibinfo{author}{\bibfnamefont{N.}~\bibnamefont{Katayama}},
  \bibinfo{author}{\bibfnamefont{E.}~\bibnamefont{Nishibori}},
  \bibinfo{author}{\bibfnamefont{H.}~\bibnamefont{Sawa}},
  \bibinfo{author}{\bibfnamefont{R.}~\bibnamefont{Ishii}},
  \bibinfo{author}{\bibfnamefont{M.}~\bibnamefont{Hagiwara}},
  \bibinfo{author}{\bibfnamefont{F.}~\bibnamefont{Bridges}},
  \bibnamefont{et~al.}, \bibinfo{journal}{Science}
  \textbf{\bibinfo{volume}{336}}, \bibinfo{pages}{559} (\bibinfo{year}{2012}).

\bibitem[{\citenamefont{Morin et~al.}(1982)\citenamefont{Morin, Schmitt, and
  du~Tremolet~de Lacheisserie}}]{PrPb3_Morin82}
\bibinfo{author}{\bibfnamefont{P.}~\bibnamefont{Morin}},
  \bibinfo{author}{\bibfnamefont{D.}~\bibnamefont{Schmitt}}, \bibnamefont{and}
  \bibinfo{author}{\bibfnamefont{E.}~\bibnamefont{du~Tremolet~de
  Lacheisserie}}, \bibinfo{journal}{J. Magn. Magn. Mater.}
  \textbf{\bibinfo{volume}{30}}, \bibinfo{pages}{257} (\bibinfo{year}{1982}).

\bibitem[{\citenamefont{Suzuki et~al.}(1997)\citenamefont{Suzuki, Kasaya,
  Miyazaki, Nemoto, and Goto}}]{PrPtBi_Suzuki97}
\bibinfo{author}{\bibfnamefont{H.}~\bibnamefont{Suzuki}},
  \bibinfo{author}{\bibfnamefont{M.}~\bibnamefont{Kasaya}},
  \bibinfo{author}{\bibfnamefont{T.}~\bibnamefont{Miyazaki}},
  \bibinfo{author}{\bibfnamefont{Y.}~\bibnamefont{Nemoto}}, \bibnamefont{and}
  \bibinfo{author}{\bibfnamefont{T.}~\bibnamefont{Goto}}, \bibinfo{journal}{J.
  Phys. Soc. Jpn.} \textbf{\bibinfo{volume}{66}}, \bibinfo{pages}{2566}
  (\bibinfo{year}{1997}).

\bibitem[{\citenamefont{Onimaru et~al.}(2005)\citenamefont{Onimaru, Sakakibara,
  Aso, Yoshizawa, Suzuki, and Takeuchi}}]{PrPb3_Onimaru05}
\bibinfo{author}{\bibfnamefont{T.}~\bibnamefont{Onimaru}},
  \bibinfo{author}{\bibfnamefont{T.}~\bibnamefont{Sakakibara}},
  \bibinfo{author}{\bibfnamefont{N.}~\bibnamefont{Aso}},
  \bibinfo{author}{\bibfnamefont{H.}~\bibnamefont{Yoshizawa}},
  \bibinfo{author}{\bibfnamefont{H.~S.} \bibnamefont{Suzuki}},
  \bibnamefont{and} \bibinfo{author}{\bibfnamefont{T.}~\bibnamefont{Takeuchi}},
  \bibinfo{journal}{Phys. Rev. Lett.} \textbf{\bibinfo{volume}{94}},
  \bibinfo{pages}{197201} (\bibinfo{year}{2005}).

\bibitem[{\citenamefont{Onimaru et~al.}(2010)\citenamefont{Onimaru, Matsumoto,
  Inoue, Umeo, Saiga, Matsushita, Tamura, Nishimoto, Ishii, Suzuki
  et~al.}}]{PrIr2Zn20_Onimaru10}
\bibinfo{author}{\bibfnamefont{T.}~\bibnamefont{Onimaru}},
  \bibinfo{author}{\bibfnamefont{K.~T.} \bibnamefont{Matsumoto}},
  \bibinfo{author}{\bibfnamefont{Y.~F.} \bibnamefont{Inoue}},
  \bibinfo{author}{\bibfnamefont{K.}~\bibnamefont{Umeo}},
  \bibinfo{author}{\bibfnamefont{Y.}~\bibnamefont{Saiga}},
  \bibinfo{author}{\bibfnamefont{Y.}~\bibnamefont{Matsushita}},
  \bibinfo{author}{\bibfnamefont{R.}~\bibnamefont{Tamura}},
  \bibinfo{author}{\bibfnamefont{K.}~\bibnamefont{Nishimoto}},
  \bibinfo{author}{\bibfnamefont{I.}~\bibnamefont{Ishii}},
  \bibinfo{author}{\bibfnamefont{T.}~\bibnamefont{Suzuki}},
  \bibnamefont{et~al.}, \bibinfo{journal}{J. Phys. Soc. Jpn.}
  \textbf{\bibinfo{volume}{79}}, \bibinfo{pages}{033704}
  (\bibinfo{year}{2010}).

\bibitem[{\citenamefont{Onimaru et~al.}({2011})\citenamefont{Onimaru,
  Matsumoto, Inoue, Umeo, Sakakibara, Karaki, Kubota, and
  Takabatake}}]{PrIr2Zn20_Onimaru11}
\bibinfo{author}{\bibfnamefont{T.}~\bibnamefont{Onimaru}},
  \bibinfo{author}{\bibfnamefont{K.~T.} \bibnamefont{Matsumoto}},
  \bibinfo{author}{\bibfnamefont{Y.~F.} \bibnamefont{Inoue}},
  \bibinfo{author}{\bibfnamefont{K.}~\bibnamefont{Umeo}},
  \bibinfo{author}{\bibfnamefont{T.}~\bibnamefont{Sakakibara}},
  \bibinfo{author}{\bibfnamefont{Y.}~\bibnamefont{Karaki}},
  \bibinfo{author}{\bibfnamefont{M.}~\bibnamefont{Kubota}}, \bibnamefont{and}
  \bibinfo{author}{\bibfnamefont{T.}~\bibnamefont{Takabatake}},
  \bibinfo{journal}{{Phys. Rev. Lett.}} \textbf{\bibinfo{volume}{{106}}},
  \bibinfo{pages}{{177001}} (\bibinfo{year}{{2011}}).

\bibitem[{\citenamefont{Cox}(1987)}]{Cox87}
\bibinfo{author}{\bibfnamefont{D.~L.} \bibnamefont{Cox}},
  \bibinfo{journal}{Phys. Rev. Lett.} \textbf{\bibinfo{volume}{59}},
  \bibinfo{pages}{1240} (\bibinfo{year}{1987}).

\bibitem[{\citenamefont{Yatskar et~al.}(1996)\citenamefont{Yatskar, Beyermann,
  Movshovich, and Canfield}}]{PrInAg2_Yaster96}
\bibinfo{author}{\bibfnamefont{A.}~\bibnamefont{Yatskar}},
  \bibinfo{author}{\bibfnamefont{W.~P.} \bibnamefont{Beyermann}},
  \bibinfo{author}{\bibfnamefont{R.}~\bibnamefont{Movshovich}},
  \bibnamefont{and} \bibinfo{author}{\bibfnamefont{P.~C.}
  \bibnamefont{Canfield}}, \bibinfo{journal}{Phys. Rev. Lett.}
  \textbf{\bibinfo{volume}{77}}, \bibinfo{pages}{3637} (\bibinfo{year}{1996}).

\bibitem[{\citenamefont{Sakai and Nakatsuji}(2011)}]{PrTr2Al20_Sakai11}
\bibinfo{author}{\bibfnamefont{A.}~\bibnamefont{Sakai}} \bibnamefont{and}
  \bibinfo{author}{\bibfnamefont{S.}~\bibnamefont{Nakatsuji}},
  \bibinfo{journal}{J. Phys. Soc. Jpn.} \textbf{\bibinfo{volume}{80}},
  \bibinfo{pages}{063701} (\bibinfo{year}{2011}).

\bibitem[{\citenamefont{Sakai et~al.}(2012)\citenamefont{Sakai, Kuga, and
  Nakatsuji}}]{PrTi2Al20_Sakai12}
\bibinfo{author}{\bibfnamefont{A.}~\bibnamefont{Sakai}},
  \bibinfo{author}{\bibfnamefont{K.}~\bibnamefont{Kuga}}, \bibnamefont{and}
  \bibinfo{author}{\bibfnamefont{S.}~\bibnamefont{Nakatsuji}},
  \bibinfo{journal}{J. Phys. Soc. Jpn.} \textbf{\bibinfo{volume}{81}},
  \bibinfo{pages}{083702} (\bibinfo{year}{2012}).

\bibitem[{\citenamefont{Matsubayashi et~al.}(2012)\citenamefont{Matsubayashi,
  Tanaka, Sakai, Nakatsuji, Kubo, and Uwatoko}}]{PrTi2Al20_Matsubayashi12}
\bibinfo{author}{\bibfnamefont{K.}~\bibnamefont{Matsubayashi}},
  \bibinfo{author}{\bibfnamefont{T.}~\bibnamefont{Tanaka}},
  \bibinfo{author}{\bibfnamefont{A.}~\bibnamefont{Sakai}},
  \bibinfo{author}{\bibfnamefont{S.}~\bibnamefont{Nakatsuji}},
  \bibinfo{author}{\bibfnamefont{Y.}~\bibnamefont{Kubo}}, \bibnamefont{and}
  \bibinfo{author}{\bibfnamefont{Y.}~\bibnamefont{Uwatoko}},
  \bibinfo{journal}{Phys. Rev. Lett.} \textbf{\bibinfo{volume}{109}},
  \bibinfo{pages}{187004} (\bibinfo{year}{2012}).

\bibitem[{\citenamefont{Tsujimoto et~al.}(2014)\citenamefont{Tsujimoto,
  Matsumoto, Tomita, Sakai, and Nakatsuji}}]{PrV2Al20_Tsujimoto14}
\bibinfo{author}{\bibfnamefont{M.}~\bibnamefont{Tsujimoto}},
  \bibinfo{author}{\bibfnamefont{Y.}~\bibnamefont{Matsumoto}},
  \bibinfo{author}{\bibfnamefont{T.}~\bibnamefont{Tomita}},
  \bibinfo{author}{\bibfnamefont{A.}~\bibnamefont{Sakai}}, \bibnamefont{and}
  \bibinfo{author}{\bibfnamefont{S.}~\bibnamefont{Nakatsuji}},
  \bibinfo{journal}{Accepted for publication in Phys. Rev. Lett.
  (arXiv:1407.0866)}  (\bibinfo{year}{2014}).

\bibitem[{\citenamefont{Matsunami et~al.}(2011)\citenamefont{Matsunami,
  Taguchi, Chainani, Eguchi, Oura, Sakai, Nakatsuji, and
  Shin}}]{PrTi2Al20_Matsunami11}
\bibinfo{author}{\bibfnamefont{M.}~\bibnamefont{Matsunami}},
  \bibinfo{author}{\bibfnamefont{M.}~\bibnamefont{Taguchi}},
  \bibinfo{author}{\bibfnamefont{A.}~\bibnamefont{Chainani}},
  \bibinfo{author}{\bibfnamefont{R.}~\bibnamefont{Eguchi}},
  \bibinfo{author}{\bibfnamefont{M.}~\bibnamefont{Oura}},
  \bibinfo{author}{\bibfnamefont{A.}~\bibnamefont{Sakai}},
  \bibinfo{author}{\bibfnamefont{S.}~\bibnamefont{Nakatsuji}},
  \bibnamefont{and} \bibinfo{author}{\bibfnamefont{S.}~\bibnamefont{Shin}},
  \bibinfo{journal}{Phys. Rev. B} \textbf{\bibinfo{volume}{84}},
  \bibinfo{pages}{193101} (\bibinfo{year}{2011}).

\bibitem[{\citenamefont{Tokunaga et~al.}(2013)\citenamefont{Tokunaga, Sakai,
  Kambe, Sakai, Nakatsuji, and Harima}}]{PrTi2Al20_Tokunaga13}
\bibinfo{author}{\bibfnamefont{Y.}~\bibnamefont{Tokunaga}},
  \bibinfo{author}{\bibfnamefont{H.}~\bibnamefont{Sakai}},
  \bibinfo{author}{\bibfnamefont{S.}~\bibnamefont{Kambe}},
  \bibinfo{author}{\bibfnamefont{A.}~\bibnamefont{Sakai}},
  \bibinfo{author}{\bibfnamefont{S.}~\bibnamefont{Nakatsuji}},
  \bibnamefont{and} \bibinfo{author}{\bibfnamefont{H.}~\bibnamefont{Harima}},
  \bibinfo{journal}{Phys. Rev. B} \textbf{\bibinfo{volume}{88}},
  \bibinfo{pages}{085124} (\bibinfo{year}{2013}).

\bibitem[{\citenamefont{Shimura et~al.}(2013)\citenamefont{Shimura, Ohta,
  Sakakibara, Sakai, and Nakatsuji}}]{PrV2Al20_Shimura13}
\bibinfo{author}{\bibfnamefont{Y.}~\bibnamefont{Shimura}},
  \bibinfo{author}{\bibfnamefont{Y.}~\bibnamefont{Ohta}},
  \bibinfo{author}{\bibfnamefont{T.}~\bibnamefont{Sakakibara}},
  \bibinfo{author}{\bibfnamefont{A.}~\bibnamefont{Sakai}}, \bibnamefont{and}
  \bibinfo{author}{\bibfnamefont{S.}~\bibnamefont{Nakatsuji}},
  \bibinfo{journal}{J. Phys. Soc. Jpn.} \textbf{\bibinfo{volume}{82}},
  \bibinfo{pages}{043705} (\bibinfo{year}{2013}).

\bibitem[{SM()}]{SM}
\bibinfo{note}{See Supplemental Material, which includes Refs. [33, 34] at [URL
  will be inserted by publisher] for the experimental method and the analysis
  of the CEF under magnetic field.}

\bibitem[{\citenamefont{Hattori and Tsunetsugu}(2014)}]{G3_Hattori14}
\bibinfo{author}{\bibfnamefont{K.}~\bibnamefont{Hattori}} \bibnamefont{and}
  \bibinfo{author}{\bibfnamefont{H.}~\bibnamefont{Tsunetsugu}},
  \bibinfo{journal}{J. Phys. Soc. Jpn.} \textbf{\bibinfo{volume}{83}},
  \bibinfo{pages}{034709} (\bibinfo{year}{2014}).

\bibitem[{\citenamefont{Araki et~al.}(2014)\citenamefont{Araki, Shimura, Kase,
  Sakakibara, Sakai, and Nakatsuji}}]{PrV2Al20_Proc_Araki14}
\bibinfo{author}{\bibfnamefont{K.}~\bibnamefont{Araki}},
  \bibinfo{author}{\bibfnamefont{Y.}~\bibnamefont{Shimura}},
  \bibinfo{author}{\bibfnamefont{N.}~\bibnamefont{Kase}},
  \bibinfo{author}{\bibfnamefont{T.}~\bibnamefont{Sakakibara}},
  \bibinfo{author}{\bibfnamefont{A.}~\bibnamefont{Sakai}}, \bibnamefont{and}
  \bibinfo{author}{\bibfnamefont{S.}~\bibnamefont{Nakatsuji}},
  \bibinfo{journal}{JPS Conf. Proc.} \textbf{\bibinfo{volume}{3}},
  \bibinfo{pages}{011093} (\bibinfo{year}{2014}).

\bibitem[{\citenamefont{Lea et~al.}(1962)\citenamefont{Lea, Leask, and
  Wolf}}]{LLW}
\bibinfo{author}{\bibfnamefont{K.}~\bibnamefont{Lea}},
  \bibinfo{author}{\bibfnamefont{M.}~\bibnamefont{Leask}}, \bibnamefont{and}
  \bibinfo{author}{\bibfnamefont{W.}~\bibnamefont{Wolf}}, \bibinfo{journal}{J.
  Phys. Chem. Solids} \textbf{\bibinfo{volume}{23}}, \bibinfo{pages}{1381}
  (\bibinfo{year}{1962}).

\bibitem[{\citenamefont{Aoki et~al.}(2002)\citenamefont{Aoki, Namiki, Matsuda,
  Abe, Sugawara, and Sato}}]{PrFe4P12_Aoki}
\bibinfo{author}{\bibfnamefont{Y.}~\bibnamefont{Aoki}},
  \bibinfo{author}{\bibfnamefont{T.}~\bibnamefont{Namiki}},
  \bibinfo{author}{\bibfnamefont{T.~D.} \bibnamefont{Matsuda}},
  \bibinfo{author}{\bibfnamefont{K.}~\bibnamefont{Abe}},
  \bibinfo{author}{\bibfnamefont{H.}~\bibnamefont{Sugawara}}, \bibnamefont{and}
  \bibinfo{author}{\bibfnamefont{H.}~\bibnamefont{Sato}},
  \bibinfo{journal}{Phys. Rev. B} \textbf{\bibinfo{volume}{65}},
  \bibinfo{pages}{064446} (\bibinfo{year}{2002}).

\end{thebibliography}

\end{document}